# Is there agreement on the prestige of scholarly book publishers in the Humanities? DELPHI over survey results

Elea Giménez-Toledo, ILIA Research Group, Institute of Philosophy (IFS), Spanish National Research Council (CSIC). Albasanz Street, 26-28, 28037, Madrid. Email: elea.gimenez@cchs.csic.es.

Jorge Mañana-Rodríguez, ILIA Research Group, Institute of Philosophy (IFS), Spanish National Research Council (CSIC). Albasanz Street, 26-28, 28037, Madrid. Email: jorge.mannana@cchs.csic.es.

**Abstract:**

Despite having an important role supporting assessment processes, criticism towards evaluation systems and the categorizations used are frequent. Considering the acceptance by the scientific community as an essential issue for using rankings or categorizations in research evaluation, the aim of this paper is testing the results of rankings of scholarly book publishers' prestige, Scholarly Publishers Indicators (SPI hereafter). SPI is a public, survey-based ranking of scholarly publishers' prestige (among other indicators). The latest version of the ranking (2014) was based on an expert consultation with a large number of respondents. In order to validate and refine the results for Humanities' fields as proposed by the assessment agencies, a Delphi technique was applied with a panel of randomly selected experts over the initial rankings. The results show an equalizing effect of the technique over the initial rankings as well as a high degree of concordance between its theoretical aim (consensus among experts) and its empirical results (summarized with Gini Index). The resulting categorization is understood as more conclusive and susceptible of being accepted by those under evaluation.

**Acknowledgements**

The authors wish to thank ANECA for supporting this study, as well as the specialists in fields of the Humanities who participated in the consultation.

**Introduction**

The existence of performance-based assessment and funding systems (Hicks 2012a; Frølich 2011) implies most of the times the need of classifying the communication channels in order to evaluate the research outputs. These kind of tools are usually to 'inform not to perform' (Sivertsen Giménez-Toledo *et al*.) in research evaluation processes. Some evaluation processes such as the Research Excellence Framework in the UK (REF, 2014) have opted not to use categorizations, classifications or rankings for the publications of researchers, using instead peer review-based procedures. Nevertheless, those are tools used in most evaluation systems in order to support informed decision making, as a guide for the evaluation and combined with expert opinion. Categorizations, classifications and rankings also serve as a mean for distinguishing scholarly journals and publishers from other types of publishers and also add value in terms of comparison, contextualizing the position of each journal or publisher.

Despite having an important role supporting assessment processes, criticism towards evaluation systems and the categorizations used are frequent. Among the best well known and numerous are those concerning the coverage and metrics of the Web of Science / Journal Citation Reports (Seglen, 1997, Jacsó, 2012, Bornmann and Marx, 2015 in example). Other information systems are not free from criticism; it is the case of ERIH (Journals under threat, 2009), Scimago Journal Rank (Mañana-Rodríguez, 2004; Jacsó, 2009) or those related to the effects of the quality label for peer reviewed books in Flanders (Borghart, 2013). Controversy is consubstantial to evaluation processes in general and to the tools developed in particular. Nevertheless, not all criticism is equally grounded on evidence: from opinion to empirical demonstration and the publication of manifestos, there is a wide diversity in the form that criticism towards scientific assessment takes.

Tools for scientific assessment should have a sound methodological basis, including transparency as a key element (Weingart, 2005); validation by experts should also be counted among the desirable features of an assessment system intended to be used responsibly and accepted by the scientific community.

After the publication in Spain of the rankings of scholarly book publishers' prestige, Scholarly Publishers Indicators (SPI hereafter), developed from the opinion of Spanish scholars and their validation by experts (Giménez-Toledo, Tejada-Artigas & Mañana-Rodríguez, 2013), the Spanish scientific assessment agency CNEAI (National Commission for the Evaluation of Research; BOE, 2014) included the rankings as a source of information and reference for the evaluation of scholarly books in Social Sciences and Humanities. At that moment, criteria used by the National Agency for Quality Assessment and Accreditation of Spain (ANECA hereafter, the agency is in charge of assessing researchers, lecturers and scholars for tenure and promotion at the national level) was being restructured together with the evaluation processes. For those two reasons, it was proposed to the research group which developed SPI to secure the results obtained in SPI using a Delphi consultation and checking the coherence between the previous and the new results (Ferrara & Bonacorsi, 2016). This contrast would then allow checking the stability of the results and serve as a proxy for their validation, thus providing further accuracy on the adequacy of their use in assessment processes. The study was done initially with publishers belonging to Humanities' fields, since in this fields the books is a key communication channel, these are the fields in which more books are published and therefore, the fields more impacted by the use of the rankings.

**Objectives**

The main objective of this research was to reach a high level of consensus among a representative sample of experts in each field of the Arts and the Humanities concerning the prestige of the publishers classified in the SPI rankings (2014). The authors consider the results of SPI (Figure 1, in example) as the starting point and the hypothesis states that a different methodology would allow the corroboration and reinforcement of prestige rankings. The operative objective was to create a refined categorization of scholarly publishers in these fields which could be used as a support for decision making in the assessment processes carried out by ANECA.

Figure 1. Prestige ranking for SPI publishers in Archaeology and Prehistory.

| \ | Spanish publishers | | | \ | Non-Spanish publishers | |
|---|---|---|---|---|---|---|
| Position | Publisher | ICEE | | Position | Publisher | ICEE |
| 1 | Ariel (Grupo Planeta) | 2.452 | | 1 | Cambridge University Press | 3.838 |
| 2 | Critica (Grupo Planeta) | 2.354 | | 2 | Oxford University Press | 2.117 |
| 3 | Akal (Akal) | 1.860 | | 3 | Routledge (Francis & Taylor Group) | 1.402 |
| 4 | Csic | 1.573 | | 4 | Archaeopress | 1.276 |
| 5 | Cátedra (Grupo Anaya, Hachette Livre) | 0.789 | | 5 | Cnrs | 1.181 |
| 6 | Ediciones Bellaterra | 0.632 | | 6 | Elsevier | 1.046 |
| 7 | Sintesis | 0.515 | | 7 | Springer | 0.714 |
| 8 | Alianza (Grupo Anaya, Hachette Livre) | 0.385 | | 8 | Oxbow Books | 0.624 |
| 9 | Aranzadi (Thomson Reuters) | 0.228 | | 9 | Blackwell | 0.597 |
| 10 | Siglo XXI De España (Akal) | 0.206 | | 10 | L'Erma Di Bretschneider | 0.446 |
| 11 | Gredos (Grupo Rba) | 0.165 | | 11 | Brepols | 0.382 |
| 12 | Universidad De Granada | 0.161 | | 12 | Academic Press (Elsevier) | 0.375 |
| 13 | Universidad Complutense De Madrid | 0.147 | | 13 | Brill | 0.281 |
| 13 | Casa De Velazquez | 0.147 | | 14 | Chicago University Press | 0.228 |
| 13 | Universitat De Barcelona | 0.147 | | 15 | Ecole Française De Rome | 0.224 |

**Note:** full data available at http://ilia.cchs.csic.es/SPI/prestigio_sectores_2014_2En.php?materia=Arqueolog%EDa%20y%20Prehistoria&tabla_esp=spi_editoriales_arqueologia_2014&tabla_extr=spi_editoriales_arqueologia_2014_extr

**Methodology**

*Technique choice:* From the various techniques which could be used for the refinement of an existing ranking (of publishers, in this case), the Delphi technique was chosen for two reasons: first, it avoids undesirable by-products of the refinement, such as multiplication of extreme values in the resulting lists due to personal interest, invisible colleges, etc. (This would plausibly happen in the case of resubmitting the lists to a large sample of experts). The list of publishers is derived from the application of the two Delphi consultations; it is a consensual list jointly developed by all participants. The only technique which has been referred to as having the desirable property of providing consensus among experts is the Delphi technique. Furthermore, it has been previously been used in the field of Social Sciences and Humanities research on assessment procedures (Hug, Ochsner and Daniel, 2013).

*Technique specifications:* Delphi technique has been extensively applied as a prospective methodology in several fields related to decision making since the 50's of the XX century (Dalkey, Brown and Cochran, 1969). The application environments are diverse and are characterized, in general, by the need of providing answers to problems in which uncertainty (concerning the achievement of the expected goals) plays an important role, thus requiring the concurrence of experts as well as the final accomplishment of a high degree of consensus among them, in order to substantiate a decision or algorithm. In a first step, its application consists on operatively defining the problem. In this case, the operative definition was provided by the existence of a previous ranking of publishers and by the objectives of the research. In a second step, experts are selected according to a sample design (see more details below). Then, a questionnaire is prepared and sent to the experts. Once the responses are gathered, these are summarized (generally by a central tendency statistic such as the mean or the median) and included in the original form of the questionnaire and sent, in a second round of consultation, to the same experts. They are, in that second round, asked to consider the attached information resulting from the summarization of the responses in the first round while providing their answer.

*Delimitation of fields:* Arts and Humanities are the fields in which more books are published from the whole set of fields in the SSH sector in Spain. According to CRUE (Spanish Universities Rectors' Conference), 66% of the scientific publications in the Arts and Humanities fields are books or book chapters. According to this data, CNEAI (Spanish National Commision for the Assessment of Research Activity, BOE 2014) and more recently ANECA (ANECA, 2016) have included specific criteria for the assessment of books in these and other fields. Considering this evidence, the methodology has been applied to the fields belonging to Arts and Humanities in the SPI scheme (Archaeology and Prehistory, Fine Arts, Arab and Hebrew Studies, Philosophy, Geography, History and Linguistics, Literature and Philology).

*Population and sample:* The population chosen was constituted by University Professors. The choice of this population was driven by three reasons. The first is the fact that, in order to become University Professor in Spanish Universities, the candidates' CV has to be evaluated by ANECA. In all cases, over 30% of the scores in that assessment process are directly linked to their publications. In fields with a large percentage of outputs in the form of books and book chapters, it implies a better chance of being not only an expert in the own field of knowledge, but also in the book culture in general as well as the publishers' prestige and reputation in the field. The second reason is the fact that University Professor is the highest position in the academic scale: a long experience as researcher and lecturer is required in order to obtain a positive result in the assessment process; thereforea good knowledge of book publishing sector in their fields is assumed. The third reason is that, since professors are not bound to assessment concerning their promotion in the academic scale, it might remove or reduce biases in their responses derived from the assumption that their responses might somehow influence further decisions concerning their academic promotion (in example, providing high scores for publishers in which they have already published).

Concerning the point in time when they have obtained their status as University Profesors, a number of years had to be chosen in order to reduce the incidence of retirement in the response pattern; in this sense, the period 2008-2015 was chosen as the time framework in which the population would be set. ANECA publishes the lists of accredited professors several times each year. The authors gathered the data from ANECA website and then searched the email addresses of all the University Profesors recognized in the 2008-2015 period. Also, the field of knowledge was gathered from the institutional websites in their universities, thus identifying the professors belonging to the fields under study.

Once the whole dataset was obtained, a stratified random sampling with proportionate allocation (to number of accredited professors in each stratum) was applied. The sample size was calculated applying the usual formula for a finite population with p=q=0.05, confidence level 95% and confidence interval=5. The following table (1) reflects the population and sample sizes required for the parameters detailed above.

**Table 1. Population size and simple size.**

| Field | Accredited professors (2008-2015) | Sample size |
|---|---|---|
| Archaeology and Prehistory | 57 | 45 |
| Fine Arts | 27 | 26 |
| Arab and Hebrew Studies | 17 | 17 |
| Philosopy | 73 | 62 |
| Geography | 80 | 67 |
| History | 377 | 191 |
| Linguistics, Literature and Philology | 481 | 214 |
| TOTAL | 1112 | 622 |

Once the sample sizes were identified for each field, a random selection of the subjects was applied. The method used for this purpose started assigning a random number to each subject (using the VBA formula 'randbetween(a;b)' in excel). The list was sorted descendent order of the random numbers. The first n subjects for each field were chosen as the random sample.

*Initial information and data transformations:* The rankings of scholarly publishers in SPI is a set of lists ordered by the so called ICEE (Quality indicator according to Experts Opinion) as the result of a large consultation among experts in SSH fields. The value of ICEE depends on the number of votes received by each publisher as well as the scores attached to each vote. It is a continuous indicator with 0 as minimum value and with an open maximum value. In order to simplify the Delphi consultation, the accumulative quartile of belonging of each publisher was calculated. In many information systems used for research evaluation purposes, the quartiles in a given distribution of, in example, Journal Impact Factor, contain a 25 % of the elements in the list of journals ordered by decreasing values of the indicator. This might be unbiased if and only if the best interpolation of the distribution is a linear equation; in that case, the likelihood of a journal being in the first quartile with regards to the likelihood of that same journal being in the second quartile would be linearly proportional to the differences in, in this case, impact factor. Nevertheless, when a distribution is highly skewed, as it is the case ICEE (and impact factor, as well), the calculation of quartiles including the same percentage of elements is strongly misleading; an element would need a much higher ICEE in order to 'fall' into the first quartile if compared with the required to fall in the second quartile than in the second compared with the third quartile. In this sense, a solution which accounts for the empirical distribution of the indicator which underlies the ranking is the accumulative form of calculation of the belonging to a quartile. In this definition of quartile, starting from a list ordered by decreasing value of ICEE, the first quartile would contain the first n publishers which added ICEE is smaller or equal to 25% of the sum of all ICEE in the distribution. The second quartile would include the journals which, after those in the first quartile, accumulate the next 25% of the sum of the ICEE in the field.

The next table (2) shows an example of how quartile distribution was calculated for Archaeology and Prehistory publishers:

**Table 2. Example of quartile calculation**

| Position | Publisher | ICEE | Accum. ICEE | Accum. % (ICEE) | Quartile |
|---|---|---|---|---|---|
| 1. | Ariel (Grupo Planeta) | 2.45 | 2.45 | 19.06 | 1 |
| 2. | Crítica (Grupo Planeta) | 2.35 | 4.81 | 37.36 | 2 |
| 3. | Akal (Akal) | 1.86 | 6.67 | 51.82 | 3 |
| 4. | Csic | 1.57 | 8.24 | 64.05 | 3 |
| 5. | Cátedra (Grupo Anaya, Hachette Livre) | 0.79 | 9.03 | 70.19 | 3 |
| 6. | Ediciones Bellaterra | 0.63 | 9.66 | 75.10 | 4 |
| 7. | Síntesis | 0.52 | 10.18 | 79.10 | 4 |
| 8. | Alianza (Grupo Anaya, Hachette Livre) | 0.39 | 10.56 | 82.10 | 4 |
| 9. | Aranzadi (Thomson Reuters) | 0.23 | 10.79 | 83.87 | 4 |
| 10. | Siglo XXI De España (Akal) | 0.21 | 10.99 | 85.47 | 4 |
| … | … | … | … | … | … |

From the distribution in quartiles of the publishers, a categorization follows in which A category is associated with any publisher belonging to the first quartile, and the homologous association occurs between the second, third and fourth quartiles and their categories, B, C and D respectively. Such as equivalence (quartiles/categories) was needed in order to tackle the Delphi consultation. For experts participating in different rounds of consultation four categories are easier to manage that n positions in the ranking.From the first round of consultation in the Delphi, the following equivalence was established between the alphabetical category and its numerical value (table 3):

**Table 3. Correspondence between quartile, alphabetical category and numerical value.**

| Quartile | Alphabetical equivalent | Numerical equivalent for further calculation |
|---|---|---|
| 1 | A | 4 |
| 2 | B | 3 |
| 3 | C | 2 |
| 4 | D | 1 |

*Delphi consultation: preparation and development.* The first online consultation was addressed to the experts in each field. The contact method was a message sent by email, looking for the commitment of the professors to participate in the two rounds of the study. A list of categorized publishers (both Spanish and non Spanish publishers) was provided to professors. Respondents were asked to a) identify those publishers well known by them b) identify -within this set- those publishers which attached category they did disagree with. The reason for this type of design is the plausible bias expectable in the case that the respondents were asked to provide a score for each of the publishers and not just for those they choose as the ones they better know. In a second page, the respondents were asked to provide a new score to the publisher.

After obtaining the responses, the average scores given to each publisher (in each field) were calculated.

In a second round (this required several reminders in order to attain the required response rate), the new scores were fed in to the questionnaire and sent to the experts.

*Response rates:* Considering that two rounds were conducted, it is relevant to distinguish between the response rates of each round. In the first round, the response rate was lower (from 68% in the case of Literature, Linguistics and Philology to 16% in the case of Geography) than in the second round, since in that moment the participants were asked answer both rounds. The response rates in the second round are high, although not in all cases 100% is reached.

The response rates in most of the fields are comparatively high for this type of consultations (Giménez-Toledo, Tejada-Artigas and Mañana-Rodríguez, 2013), but in the case of Philosophy and Geography these rates were too low as to consider their responses representative. For this reason the consultation rounds with these two fields were repeated, choosing a new randomly sample to which the questionnaires were sent. The population in this case consisted ofthe experts belonging to the assessment panels and committees in the evaluation program ACADEMIA (for the assessment of lecturers with regards to the figures of Professor and tenured lecturer) and the experts belonging to CNEAI (National Commission for the assessment of Research Activity, which mission is the evaluation of six-year periods of research activity and which positive evaluation is accompanied by salary incentives) in the areas 11 (Philosophy, Philology and Linguistics), 8 (Economics and Bussiness) and 10 (History, Geography and Arts). Although the population in this case was lower than in the first attempt, experts from the panels might be much more motivated for this kind of surveys due to their involvement in research evaluation.In order to identify the population, publicly available information regarding the composition of the panels in 2015 was queried, while in the case of CNEAI the homologous documentation was revised, in this case in the period 2004-2015 with the exception of 2007 for which there is no publicly available data. The reason for choosing a different time window for each documentation is the fact that the number subjects in each ANECA programs is much larger than in the case of CNEAI, thus making necessary a widening of the time framework in the same case as to reach a population N sufficient for the needs of a sample size which could compensate the insufficient response rate in the initial round. The questionnaire in this specifically targeted extension of the sample was sent to 29 Philosophy professors and 13 Geography professors.

The response rates obtained are the following (table 4):

**Table 4. Response rates for both rounds**

| | FIRST ROUND | | | SECOND ROUND | | |
|---|---|---|---|---|---|---|
| Field | Sample n | Number of answers | Response rate | Sample n | Number of responses | Response rate |
| Archaeology and Prehistory | 45 | 16 | 35.56 | 16 | 16 | 100 |
| Fine Arts | 26 | 6 | 23.08 | 6 | 6 | 100 |
| Arab and Hebrew studies | 17 | 4 | 23.53 | 4 | 4 | 100 |
| Philosophy | 91(62+29) | 20(14+6) | 21.98 | 20 | 18 | 90 |
| Geography | 80(67+13) | 13(8+5) | 16.25 | 13 | 12 | 92.3 |
| History | 191 | 97 | 50.79 | 97 | 81 | 83.51 |
| Linguistics, Literature and Philology | 215 | 145 | 67.76 | 145 | 117 | 80.6 |
| TOTAL. ALL FIELDS | 664 | 301 | 45.33 | 301 | 254 | 92.36 |

**Note:** the response rate of the second round has been calculated on the number of subjects who responded to the first round. In parenthesis is the sample size, number of subjects and responses in the extension phase of the questionnaire for the fields of Philosophy and Geography).

In order to better understand the whole process, we present here a specific example for Springer in the case of History. The initial position of the publisher in SPI was 2 (third quartile in the distribution of the ICEE indicator). In the first round, Springer was mentioned by 13 experts in History, whose responses were: (4;3;2;2;2;2;2;2;3;4;3;3;3), $\bar{x}=2,73$, which nearest whole number is 3. In the second round, the specialists knew that the position of Springer as a result of the first round was 3 and it was voted by 9 experts whose scores were (4;4;3;3;3;3;3;2;2). The average resulting from this second round was 3. The average between 2.73 (the value in the first round) and 3 (the value in the second round) is 2,875, which nearest whole number is 3. 3 is the final result in the case of this publisher (for the field of History; the publisher would have different scores in those field in which it has been voted).

**Results:**

The following descriptive statistics describe the results obtained after the application of the technique to the set of publishers by field (Spanish and non Spanish publishers, tables 5 and 6 respectively). The mean is calculated on the difference between the new positions and the previous positions (in both cases ranging from 1 to 4, and considering the new position minus the previous position).

**Table 5. Descriptive statistics of the position changes between the first and last position of Spanish publishers.**

| SPANISH PUBLISHERS | Archaeology and Prehistory | Fine arts | Arab and Hebrew studies | History | Linguistics, Literature and Philology | Geography | Philosophy |
|---|---|---|---|---|---|---|---|
| MEAN | 0,78 | 0,47 | 0,31 | 0,43 | 0,71 | 1.02 | 1.01 |
| SD | 0,35 | 0,50 | 0,43 | 0,49 | 0,41 | 0.45 | 0.39 |

**Table 6. Descriptive statistics of the position changes between the first and last position of non-Spanish publishers.**

| NON-SPANISH PUBLISHERS | Archaeology and Prehistory | Fine arts | Arab and Hebrew studies | History | Linguistics, Literature and Philology | Geography | Philosophy |
|---|---|---|---|---|---|---|---|
| MEAN | 1,11 | 0,89 | 0,73 | 1,05 | 1,08 | 0.89 | 1.18 |
| SD | 0,75 | 0,81 | 0,71 | 0,58 | 0,45 | 0.40 | 0.50 |

In all cases, the mean change from the previous values to the new ones shows a positive trend (i.e., higher categories for publishers), stronger in the non-Spanish publishers than in the Spanish publishers. The standard deviations are, nevertheless, strong in some cases such as Fine Arts (Spanish publishers), which points towards a large diversity of differences in the response, but which might also be affected by the reduced number of publishers in the list of this field.

As a field-specific example, the following chart plots the observed tendency towards a higher scoring in the second round as well as reduced concentration of prestige for the field of Archaeology and Prehistory. This result can be observed in all analyzed fields. The following chart (1) shows the distribution of positions for Spanish publishers in Archaeology and Prehistory.

**Chart 1. Distribution of former and latter positions given to Spanish publishers in Archaeology and Prehistory.**

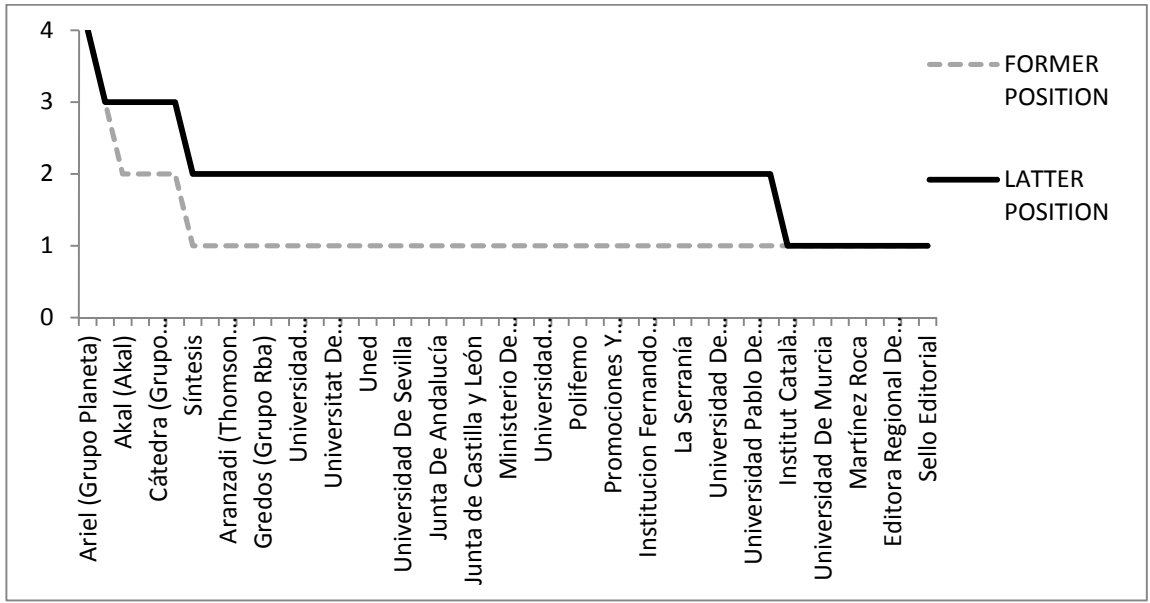

**Note**: only 25 publishers are included in this chart. See the complete chart at https://public.tableau.com/views/ArchaeologyandPrehistory_BeforeandafterDelphitechnique_/Hoja1?:embed=y&:display_count=yes

Concerning concentration indices, Gini Index show lower concentration levels / higher inequality levels in the results of the second round when compared to the initial dataset. The following chart (Chart 2) reflects the Lorentz Curve for both situations (before and after the application of the Delphi technique) for the field of Archaeology and Prehistory.

**Chart 2. Lorenz curves and Gini Index for the distribution of publishers' positions before and after the application for the Delphi technique (Spanish publishers in Archaeology and Prehistory).**

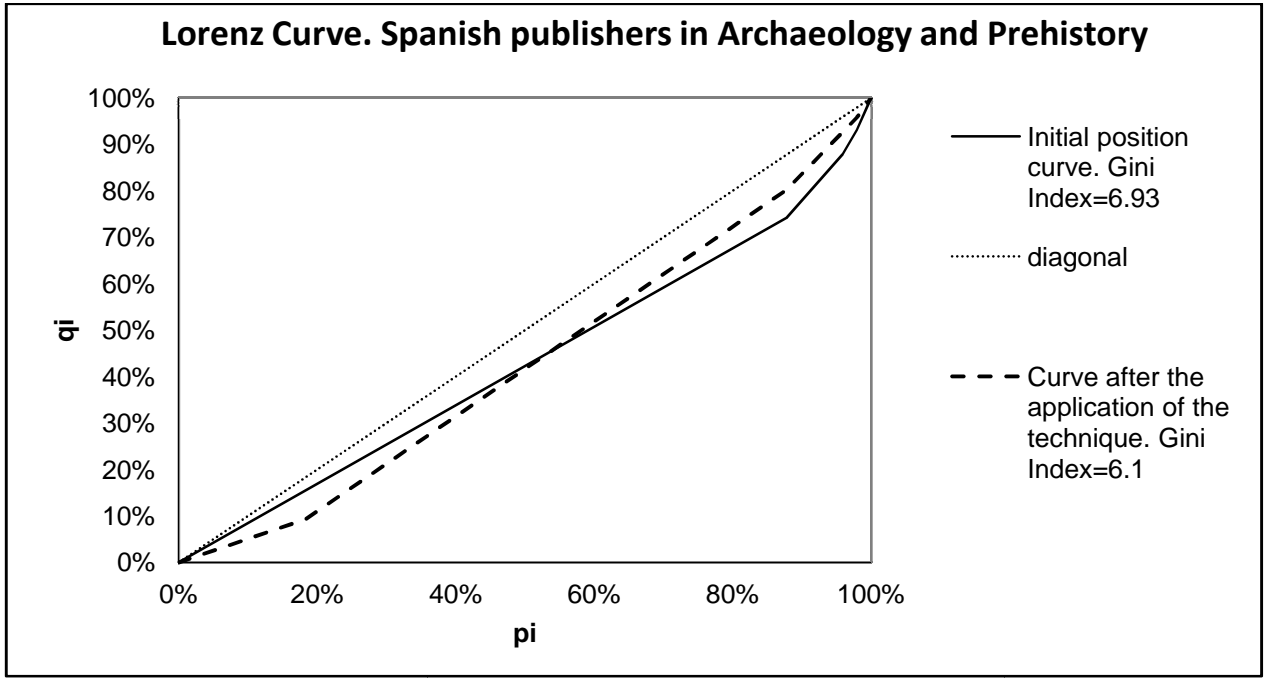

**Conclusions**

The successive rounds using the Delphi technique have yielded a reasonably good response rate, although it does not met the desired n for the parameters initially established. It is important to mention here that, despite the lack of full sample size and the consequent statistical variation of the parameters derived from it, the random sampling together with the existence of a large number of experts makes this application of the technique different (and, as we understand it, better suited for the generalization and reliability of the results) from the usual methodological approach. In its most frequent form, it usually relies on a selection of experts by the researcher, this involving also the determination of their number from the available possibilities, thus leaving more variability in the responses potentially imputable to the choice of the researcher than to the randomness and its desirable equalizing influence in the potential response biases present in any survey-based study. Considering that the results of a Delphi technique are generally considered sufficient and reliable with an intentional selection of the experts and with an arbitrary selection of their number, the lack of completeness of the response rate concerning the sample size is not understood here as a major flaw of the research carried out, although it is a desirable objective.

The results also shed light on a criticism informally exerted on the starting point information (mainly from Spanish publishers which do not appear in the ranking or do so with low values). Since the over 2700 researchers who participated in the survey which produced the initial results and rankings are subject to assessment, it has been pointed out that some of them might choose to give high scores to the publishers in which they have published in order to improve their chances of obtaining a positive assessment by agencies (the participants in the first survey were tenured lecturers as well as professors). Nevertheless, in this research, only professors were asked; they are not subject to assessment by ANECA since they reached the top position in academia professional categories and many of them are not subject to assessment by CNEAI either. With this sample, almost all of the reasons for a response pattern biased towards a better result in their particular research assessment exercises are removed. In that circumstance, if we assume that the first results were biased, strong differences in the set of publishes resulting from the consultation (both in composition and position) would be observed. This is not the case: even when given the chance to point out other publishers, the respondents did not do so in any case, and the trend in the position change is positive and without strong differences with regards to the previous ones. From this observation on the nature of the sample used in this study it can be concluded that the level of response bias in the initial consultation which yielded the Scholarly Publishers Indicators is low.

Also, it can be concluded that there is a high degree of concordance between the theoretical objective of the technique (Consensus among experts) and its empirical materialization (a significant reduction of the inequality between the initial and final distribution of scores), as provided by the difference in the curves or Gini Index.

These results are aligned with the conclusions obtained in the study by Ferrara and Bonaccorsi (2016). They worked on one of the key issues for the assessment of Social Sciences and the Humanities: the translation of qualitative judgments of experts into quantitative indicators or, in other words, in tools leading to an increase in the efficiency of the overloaded processes of scientific evaluation. The analyses had the virtue of being supported by a large amount of data gathered from real assessment processes, which makes conclusions highly clarifying. The proposed method shows that qualitative judgments and categorizations are not as far from each other as sometime it is stressed out. In our study, we can conclude something similar since the SPI rankings are the translation of the experts' opinion in quantitative indicators and categories, and these categories seem to be supported by other experts using other methodology. The results point out towards a generalized knowledge of the publishers already included in the initial prestige rankings by the respondents. This allows to affirm that this study shows evidence of validation of the previous results shown in SPI and, through a 'refinement' via Delphi, it is possible to qualify them, also making it easier to present the results in the form of a **categorization** which acceptance is likely to be higher since it does not provide individual positions for each publisher (as it is the case in the ranking). Also, it can be concluded that there is a high degree of concordance between the theoretical objective of the technique (consensus among experts) and its empirical materialization (a significant reduction of the inequality between the initial and final distribution of scores), as provided by the difference in the curves or Gini Index.

The most relevant conclusion of this study is the fact that, after the two consultations, there is a consensual higher qualification of all publishers and a reduced polarization of the scores in the initial positions of the ranking.